\documentclass[aps,twocolumn,superscriptaddress,nofootinbib]{revtex4-2}
\usepackage{dcolumn}
\usepackage{bm}
\usepackage{graphicx}
\usepackage{amsfonts}
\usepackage{bbm}
\usepackage{dsfont}
\usepackage{float}
\usepackage{graphicx}
\usepackage{bm}
\usepackage{amssymb}
\usepackage{multibib}
\usepackage{color}
\usepackage{enumerate}
\usepackage{amsmath}
\usepackage{amstext}
\usepackage{latexsym}
\usepackage{braket}
\usepackage{appendix}
\usepackage{textcomp}
\usepackage[usenames,dvipsnames]{xcolor}
\usepackage[colorlinks=true,citecolor=Blue,linkcolor=RubineRed,urlcolor=Blue]{hyperref}

\begin{document}

\title{Probing quantum chaos in multipartite systems}
\date{\today}
\author{Zan Cao}
\affiliation{School of Physical Science and Technology, Soochow University, Suzhou 215006, China}
\author{Zhenyu Xu \href{https://orcid.org/0000-0003-1049-6700}{\includegraphics[scale=0.05]{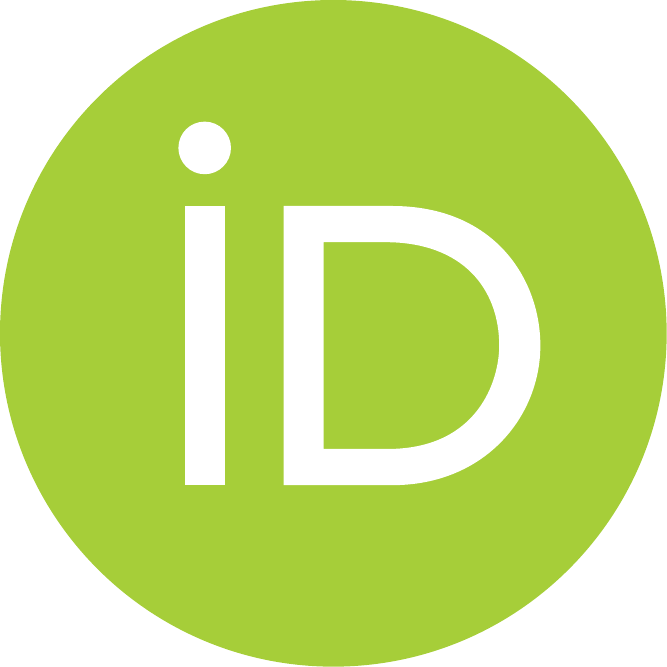}}}
\email{zhenyuxu@suda.edu.cn}
\affiliation{School of Physical Science and Technology, Soochow University, Suzhou 215006, China}
\author{Adolfo del Campo \href{https://orcid.org/0000-0003-2219-2851}{\includegraphics[scale=0.05]{orcidid.pdf}}}
\email{adolfo.delcampo@uni.lu}
\affiliation{Department of Physics and Materials Science, University of Luxembourg, L-1511 Luxembourg, Luxembourg}
\affiliation{Donostia International Physics Center, E-20018 San Sebasti\'an, Spain}

\begin{abstract}
Understanding the emergence of quantum chaos in multipartite systems is
challenging in the presence of interactions. We show that the contribution
of the subsystems to the global behavior can be revealed by probing the full
counting statistics of the local, total, and interaction energies. As in the
spectral form factor, signatures of quantum chaos in the time domain dictate
a dip-ramp-plateau structure in the characteristic function, i.e., the
Fourier transform of the eigenvalue distribution. With this approach, we
explore the fate of chaos in interacting subsystems that are locally
maximally chaotic. Global quantum chaos can be suppressed at strong
coupling, as illustrated with coupled copies of random-matrix Hamiltonians
and of the Sachdev-Ye-Kitaev model. Our method is amenable to experimental
implementation using single-qubit interferometry.
\end{abstract}

\maketitle

\section{Introduction}

Identifying signatures of chaos in the quantum domain remains a nontrivial
task in complex systems \cite{Haake}. Quantum chaos has manifold
applications and appears in different fields involving the study of
many-body complex quantum systems \cite{MethaBook,Zelevinsky96,Guhr98},
statistical mechanics of isolated quantum systems \cite{Srednicki94,DAlessio16}, anti-de Sitter black holes \cite{Hayden2007,Sekino2008,Stephen2014,Maldacena2016,Cotler2017}, holographic
quantum matter \cite{Franz2018}, and quantum information science \cite{Forrester10,RN1218}, among other examples. Several diagnostic tools for
quantum chaos have been proposed. They include the spectral form factor
(SFF) \cite{Haake}, fidelity decay in short-time \cite{Emerson2002PRL} and long-time regimes \cite{Leonski2009PLA}, Loschmidt echo (LE) \cite{Gorin06}, out-of-time-order correlator (OTOC) \cite{OTOC1969}, quantum circuit complexity \cite{Susskind2016,RN954}, etc. Connections among
these diagnostics have been explored in specific areas ranging from
many-body systems to quantum field theory \cite{Gorin06,Roberts2017,Chenu2018,ChaosWeb,Yan2020PRL,bhattacharyya2021web}.
More recently, diagnostics of quantum chaos have been extended to open
systems to account for the effect of decoherence and dissipation \cite{Karkuszewski02,Poulin2004PRL,Syzranov2018,Xu19,delcampo19,Jan2019,Yoshida2019,Lucas2019,Zhenyu2021PRB,Zanardi2021,Anand21,Julien2021}.

\begin{figure}[t]
\centering
\includegraphics[width=0.8\linewidth]{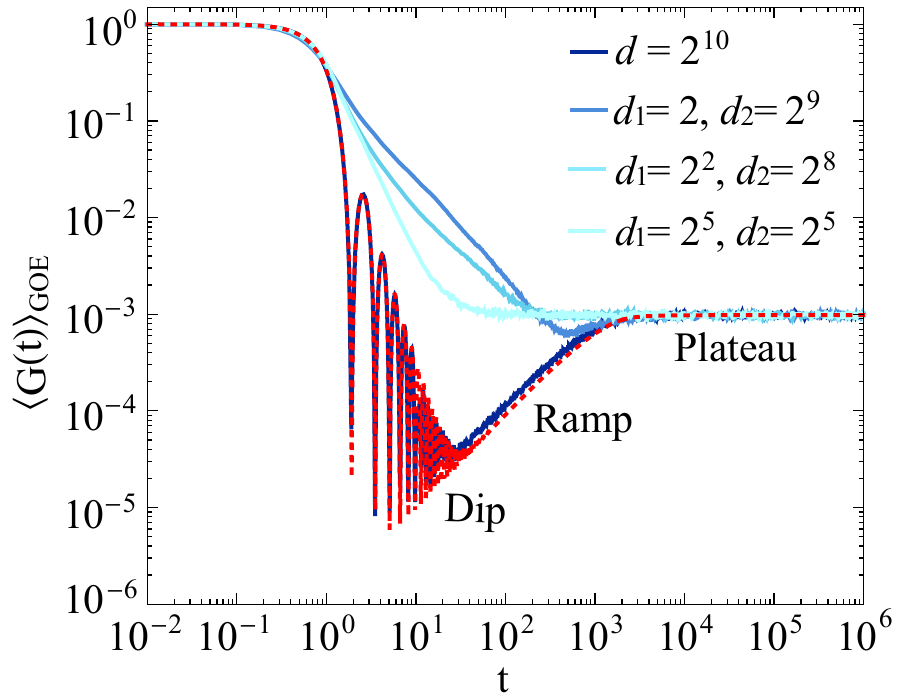}
\caption{\textbf{The dip-ramp-plateau structure: an indicator of quantum chaoticity.} The absolute square value of the generating function $G(t)$ for the total energy distribution averaged over $\mathrm{GOE}(d)$ ($d=2^{10}$) and $\mathrm{GOE}(d_{1})\otimes \mathrm{GOE}(d_{2})$ ($d_{1}d_{2}=d$), with 500 independent realizations, and $\beta =0.01$. Analytical Eq. (\ref{FGOE}) for $\mathrm{GOE}(d)$ is depicted in red color. For systems with less chaoticity than the full \textrm{GOE}, the span of the ramp will shrink or even vanish.}
\label{fig1}
\end{figure}

A prominent signature of quantum chaos is the repulsion among energy levels.
For instance, the spacing between nearest-neighbor levels follows the
Wigner-Dyson distribution in quantum chaotic systems, while it is described
by Poisson statistics in the presence of conserved quantities (e.g., in
integrable systems) \cite{Borgonovi16}. The SFF is proportional to $|Z(\beta +it)/Z(\beta )|^{2}$, where $Z(\cdot $) is the partition function and $\beta =1/k_{B}T$. This quantity probes the level statistics of both close and far-separated energy eigenvalues, providing a tool to detect the ergodic nature of the system \cite{Haake}. For a generic chaotic system, the SFF exhibits a dip-ramp-plateau structure [see e.g., Fig. \ref{fig1}]. Its short-time decay forms a slope. The physical origin of the subsequent ramp is the long-range repulsion between
energy levels \cite{Cotler2017}. The transition from the slope to the ramp
forms the dip. The final plateau originates from the finite Hilbert space
dimension and approaches a constant value $Z(2\beta )$/$Z(\beta )^{2}$ in the absence of degeneracies in the energy spectrum. The SFF has been widely employed in the study of the discrete energy spectrum of quantum chaotic systems \cite{Leviandier86,WilkieBrumer91,Alhassid93,Ma95,Cotler2017,Dyer2017,delcampo17,Zoller2022PRX}.

Quantum chaotic systems composed of multipartite subsystems subject to generic
interactions typically have a complicated energy spectrum \citep{Lakshminarayan2016PRL,Lakshminarayan2018PRB,Styliaris21,Zoller2022PRX}. We shall focus on a global subsystem composed of strongly-chaotic subsystems, interacting with
each other. In this setting, any subsystem can be seen as an open quantum
system embedded in an environment, composed of the remaining subsystems. The
subsystem dynamics is thus governed by dissipative quantum chaos \cite{Haake}, which is currently under exhaustive study \cite{Xu19,Can19,Denisov19,delcampo19,Lucas2019,Sa2020SpecTrans,Zhenyu2021PRB,Julien2021,Tarnowski21,Sa2021lindbladian}. We shall depart from the standard practice of assuming an effective open quantum dynamics, as ubiquitously done in the literature. Instead, we will
account for the exact unitary dynamics of the global composite system, with
no approximations (e.g., without invoking the Markovian description or an
effective master equation). The above diagnostics can be employed to detect
global quantum chaos in multipartite systems \cite{Styliaris21}. However, apart from proposals like the fidelity-based SFF \cite{Zhenyu2021PRB,Julien2021} and the related partial SFF \citep{Zoller2022PRX}, they are not suited to directly detect how chaotic behavior stems from the subsystems and their interactions. In this work, we provide an experimentally realizable approach to this end by considering the measurement of an energy observable $X$, which can be the Hamiltonian of a subsystem or the
interaction energy. As measurement outcomes are stochastic, we propose to
study the full counting statistics, characterized by the eigenvalue
distribution of $X$ at thermal equilibrium. Its Fourier transform, the
characteristic function, reveals chaotic behavior through the
dip-ramp-plateau structure. Its analysis shows that strong interactions
among the different subsystems can suppress the global chaotic behavior of
the multipartite system, even when the subsystems are maximally chaotic, as
revealed by the study of global and local observables. This scheme not only
provides a convenient theoretical tool to diagnose quantum chaos in complex
multipartite quantum systems, but it can be experimentally realized by using
single-qubit interferometry with an ancillary qubit.

Our paper is organized as follows. In Sec. II, we introduce the
characteristic function of an energy observable, which is then used
as a tool to detect the chaotic features stemming from the subsystems and their
interactions in Sec. III. Then, we employ this method to analyze the chaotic
behavior in multipartite systems sampled from the Gaussian orthogonal
ensemble ($\mathrm{GOE}$) in Random Matrix Theory (RMT) \cite{MethaBook,VivoBook} in Sec. IV and the coupled Sachdev--Ye--Kitaev (cSYK)
models \cite{maldacena2018eternal,Zhai2017,Song2017,Yao2017,Garica2019,Plugge2020,Qi2020,Sahoo2020,Haenel2021,Zhai2021}
in Sec. V. Finally, we summarize in Sec. VI with concluding remarks and a
brief discussion of potential applications.

\section{The characteristic function of the energy observable $X$}

Let $\mathcal{H}=\bigotimes_{l}\mathcal{H}_{l}$ be the Hilbert space of a
multipartite system and $X$ be an energy observable of a local subsystem in
the subspace $\bigotimes_{k}\mathcal{H}_{k}$ $\subseteq \mathcal{H}$. We
focus on the Hamiltonian of the subsystems (local energy) and the
interaction energy, as choices of the observable $X$. The probability
distribution of the observable $X$ with eigenvalues $\{x\}$ averaged over an
initial thermal equilibrium state $\rho _{\mathrm{th}}=e^{-\beta H}/Z(\beta) $ is \cite{Zhenyu2019PRLmanybody}
\begin{equation}
P(x)=\mathrm{tr}\left[ \rho _{\mathrm{th}}\delta (X-x)\right] .
\end{equation}%
The eigenvalue probability distribution $P(x)$ encodes the full counting
statistics of the observable $X$, that is, the probability to find the
system in an eigenstate with eigenvalue $x$ when prepared in the state $\rho
_{\mathrm{th}}$. In terms of the integral representation of the Dirac delta
function, the probability distribution can be expressed as the Fourier
transform
\begin{equation}
P(x)=\frac{1}{2\pi }\int_{-\infty }^{+\infty }g(t)e^{-itx}dt,
\end{equation}%
of the characteristic (moment generating) function
\begin{equation}
g(t)=\mathrm{tr}\left( \rho _{\mathrm{th}}e^{itX}\right) ,  \label{Cf}
\end{equation}%
that captures the statistical properties of the spectrum of the observable $%
X $. While in the following we refer to $t$ as a time variable, it is to be
understood as the Fourier conjugate to $x$.

Experimentally, the characteristic function Eq. (\ref{Cf}) can be measured
by introducing an auxiliary qubit coupled to the system. This technique
known as single-qubit interferometry has been widely used in measuring fidelity decay \cite{Poulin2004PRL}, form factors of Floquet operators \cite{Poulin2003PRA}, local density of states \cite{Emerson2004PRA}, LE \cite{Quan2006,Zhang2008}, work statistics \cite{Dorner2013,Mazzola2013,Tiago2014}, Lee-Yang zeros \cite{Wei12,Peng15,RN1221}, OTOC \cite{Swingle2016}, quantum-state reconstruction of dark systems \cite{Cai2019}, full distribution of many-body observables \cite{Zhenyu2019PRLmanybody}, and SFF \cite{Zoller2020PRXQ}. The key procedure is
to perform a controlled $X$ gate conditioning on the auxiliary qubit, i.e.,
\begin{equation}
U(t)=\left\vert 1\right\rangle \left\langle 1\right\vert \otimes \exp
(itX)+\left\vert 0\right\rangle \left\langle 0\right\vert \otimes \openone.
\end{equation}%
This allows us to recover the real and imaginary parts of Eq. (\ref{Cf}) by
measuring a pair of Pauli operators on the ancillary qubit \cite%
{Zhenyu2019PRLmanybody}, i.e., \textrm{Re}$(g(t))=$\textrm{tr}$(\sigma
_{z}\rho _{\mathrm{ancillary}})$ and \textrm{Im}$(g(t))=$\textrm{tr}$(\sigma
_{y}\rho _{\mathrm{ancillary}})$, where $\rho _{\mathrm{ancillary}}=\mathrm{%
tr}_{\mathrm{system}}(\mathrm{H}\otimes \openone U(t)\left\vert
+\right\rangle \left\langle +\right\vert \otimes \rho _{\mathrm{th}}U^{\dag
}(t)\mathrm{H}\otimes \openone)$, $\mathrm{H}$ is the Hadamard gate, and $%
\left\vert +\right\rangle =(\left\vert 0\right\rangle +\left\vert
1\right\rangle )/\sqrt{2}$.

\begin{figure*}[!]
\centering
\includegraphics[width=0.85\linewidth]{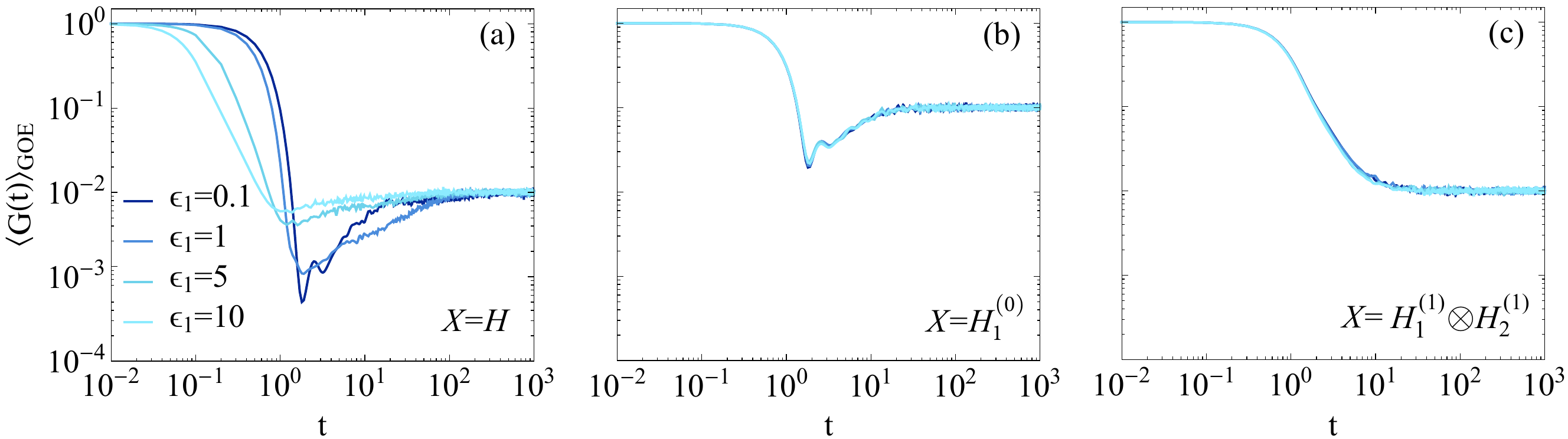}
\caption{\textbf{Chaos is suppressed by enhancing the interaction in a bipartite system.} (a) Equation (\protect\ref{CF}) averaged over \textrm{GOEs }for observables $X=H$ ($H=H_{1}^{(0)}+H_{2}^{(0)}+\protect \epsilon _{1}H_{1}^{(1)}\otimes H_{2}^{(1)}$) with coupling strength $\protect\epsilon _{1}=0.1,$ $1,$ $5,$ and $10$ respectively. $H_{1}^{(0,1)}$ and $H_{2}^{(0,1)}$ are sampled from \textrm{GOE} independently, with 500 realizations, $d_{1}=d_{2}=10$, and $\protect\beta =0.01$ (for other temperature and number of realizations, see in Appendix \protect\ref{appB}).
The size of dip-ramp shrinks when the coupling constant $\protect\epsilon_{1}$ is enhancing, implying that chaos is suppressed. The \textrm{GOE } averaged $G(t) $ is depicted with $X=H_{1}^{(0)}$ in (b) and $X=H_{1}^{(1)}\otimes H_{2}^{(1)}$ in (c), respectively, in which $H_{1}^{(1)}\otimes H_{2}^{(1)}$ with no obvious dip-ramp-plateau structure plays an important role in attenuating the chaoticity.}
\label{fig2}
\end{figure*}

\section{A probe for quantum chaos in multipartite systems}

In what follows, we consider the absolute square value of the generating
function in Eq. (\ref{Cf}), i.e.,
\begin{equation}
G(t)=|g(t)|^{2}=|\mathrm{tr}\left( \rho _{\mathrm{th}}e^{itX}\right) |^{2},%
\text{ }t\in \lbrack 0,\infty ),  \label{CF}
\end{equation}%
as a tool for probing quantum chaos in multipartite systems, identifying
contributions from the subsystems and their interactions. The choice $%
X=H_{j} $ represents the local energy that can be employed to diagnose the
chaotic behavior contributed by the $j$th subsystem in an $N$-partite
system. By contrast, the observable $X=H_{j}\otimes H_{k}\otimes H_{l}$ can
be used to detect signature of quantum chaos attributed to the interactions
among subsystems $j$, $k$, and $l$.

The original motivation behind Eq. (\ref{CF}) is based on the fact that when
the observable $X$ is chosen as the global Hamiltonian $X=H$ $\left( H\in
\mathcal{H}\right) $, the absolute square value of the generating function
in Eq. (\ref{CF}) turns out to be $G(t)=\left\vert \frac{Z(\beta +it)}{%
Z(\beta )}\right\vert ^{2}$, which equals the SFF \cite%
{Papadodimas15,Cotler2017,ChaosWeb} and can be interpreted as the fidelity
between the initial coherent Gibbs state (or a thermofield double state) and
the state resulting from its evolution \cite{delcampo17,Zhenyu2021PRB}. In
addition, when $X$ is a small perturbation of the Hamiltonian $H$, i.e., $%
H=H_{0}+X$, and commutes with $H$ (or $H_{0}$), Eq. (\ref{CF}) is similar to the LE $G(t)=\left\vert \left\langle \psi
_{0}\right\vert e^{itH}e^{-itH_{0}}\left\vert \psi _{0}\right\rangle
\right\vert ^{2}$, which captures the overlap between two identical initial
states $ (\left\vert \psi _{0}\right\rangle) $ evolving under slightly different Hamiltonians $H$
and $H_{0}$ \cite{Gorin06}. Note that according to the \
Baker--Campbell--Hausdorff formula, Eq. (\ref{CF}) and the LE differ in the
general case, when $[X,H]\neq 0$.

We emphasize that the observable $X$ in Eq. (\ref{CF}) is not required to
represent a small perturbation or to commute with $H$. When it describes the
local energy of a subsystem $\bigotimes_{k}\mathcal{H}_{k}$ $\subseteq
\mathcal{H}$, Eq. (\ref{CF}) provides the possibility to directly detect the
chaotic behavior contributed by the subsystems or interactions to the global
multipartite system. It can be used to either diagnose quantum chaos of a
one-partite system (as done by the SFF) or a structured multipartite system.
To support this observation, we illustrate its use in the following examples
involving coupled random-matrix Hamiltonians and the coupled SYK model.

\section{Probing the chaoticity in coupled random-matrix Hamiltonians}

Consider a $N-$partite system with a general Hamiltonian of the form
\begin{equation}
H=\sum_{j=1}^{N}H_{j}^{(0)}+\epsilon _{1}\sum_{j<k=2}^{N}H_{j}^{(1)}\otimes
H_{k}^{(1)}+\cdots +\epsilon _{N-1}\bigotimes_{j=1}^{N}H_{j}^{(N-1)},
\end{equation}%
where $H_{j}^{(0)}$ is the Hamiltonian of the $j$th subsystem and $\epsilon
_{l-1}$ is the coupling constant for the $l$-partite interaction $%
\bigotimes_{j=1}^{l}H_{j}^{(l-1)}$.

For the sake of illustration, let us sample the Hamiltonians from the
Gaussian orthogonal ensemble ($\mathrm{GOE}$) \cite{MethaBook,VivoBook},
which is a paradigmatic random matrix ensemble for physical applications
involving systems with time-reversal symmetry and exhibiting quantum chaos.
\textrm{GOE} is the ensemble of real symmetric matrices, whose elements are
chosen at random from a Gaussian distribution. The joint probability density
of $H_{j}\in \mathrm{GOE(}d_{j}\mathrm{)}$ ($d_{j}$ denotes the dimension of
the Hilbert space $\mathcal{H}_{j}$) is proportional to $\exp (-\frac{1}{%
2\sigma ^{2}}\mathrm{tr}H_{j}^{2})$, where $\sigma $ is the standard
deviation of the random matrix elements of $H_{j}$.

The first example we consider is $N=1$ and $X=H$. In this scenario, we focus
on $H\in \mathrm{GOE(}d\mathrm{)}$ and $H\in \mathrm{GOE(}d_{1}\mathrm{)}%
\otimes \mathrm{GOE(}d_{2}\mathrm{)}$ ($d_{1}d_{2}=d$) as an example.
Equation (\ref{CF}) averaged over the full \textrm{GOEs [i.e., GOE(d)] }%
yields
\begin{equation}
\left\langle G(t)\right\rangle _{\mathrm{GOE}}\doteq \frac{\left\langle
Z(2\beta )\right\rangle _{\mathrm{GOE}}\cdot C_{\mathrm{GOE}}+\left\vert
\left\langle Z(\beta +it)\right\rangle _{\mathrm{GOE}}\right\vert ^{2}}{%
\left\langle Z(\beta )\right\rangle _{\mathrm{GOE}}^{2}},  \label{FGOE}
\end{equation}%
where $\left\langle \cdot \right\rangle $ represents the
ensemble average and $\doteq $ denotes the annealing approximation \cite%
{Cotler2017}. In Eq. (\ref{FGOE}), the \textrm{GOE }averaged partition
function is given by (see Appendix \ref{appA})

\begin{equation}
\left\langle Z(x)\right\rangle _{\mathrm{GOE}}=\frac{\sqrt{d}\mathrm{I}%
_{1}(2\sigma \sqrt{d}x)}{\sigma x},  \label{par-GOE}
\end{equation}%
where $\mathrm{I}_{n}(\cdot )$ is the modified Bessel function of first kind
and order $n$, and the coefficient reads

\begin{equation}
C_{\mathrm{GOE}}=\left\{
\begin{array}{ll}
\frac{t\sigma }{\sqrt{d}}-\frac{t\sigma }{2\sqrt{d}}\ln \left( 1+\frac{%
t\sigma }{\sqrt{d}}\right) , & t\leq 2\sqrt{d}/\sigma , \\
2-\frac{t\sigma }{2\sqrt{d}}\ln \frac{t+\sqrt{d}/\sigma }{t-\sqrt{d}/\sigma }%
. & t>2\sqrt{d}/\sigma .%
\end{array}%
\right.  \label{CGOE}
\end{equation}

As shown by the red dotted curve (or the dark blue curve by numerical
simulations) in Fig. \ref{fig1}, Eq. (\ref{FGOE}) exhibits a typical feature
of quantum chaos, namely a dip-ramp-plateau structure. The early decay from
unit value comes to an end, forming a dip (also known as correlation hole)
with the onset of a ramp. The latter extends until it saturates at a plateau
value at the characteristic plateau time $t_{\mathrm{plateau}}=2\sqrt{d}%
/\sigma $ [see Eq. (\ref{CGOE})]. The existence of the ramp, a period of
linear growth of $\left\langle G(t)\right\rangle _{\mathrm{GOE}}$, is a
consequence of the repulsion between long-range energy levels \cite%
{Cotler2017}. This long-range repulsion causes the energy levels to be
anticorrelated. The plateau stems from the discrete energy spectrum, whose
height is $\left\langle Z(2\beta )\right\rangle _{\mathrm{GOE}}/\left\langle
Z(\beta )\right\rangle _{\mathrm{GOE}}^{2}$. Similar chaotic features have
been studied in the Gaussian Unitary Ensemble (\textrm{GUE}) \cite%
{delcampo17,Chenu2019,Zhenyu2021PRB}, Gaussian ensembles\textrm{\ }under%
\textrm{\ }infinite temperature \cite{JunyuLiu2018}, and Sachdev--Ye--Kitaev
(SYK) models \cite{Cotler2017,Liu2017,Liujunyu2018}.

For systems with less chaoticity than the full GOE, the span
of the ramp will shrink or even disappear (see light blue curves in Fig. \ref{fig1}).

\begin{figure}[!]
\centering
\includegraphics[width=0.99\linewidth]{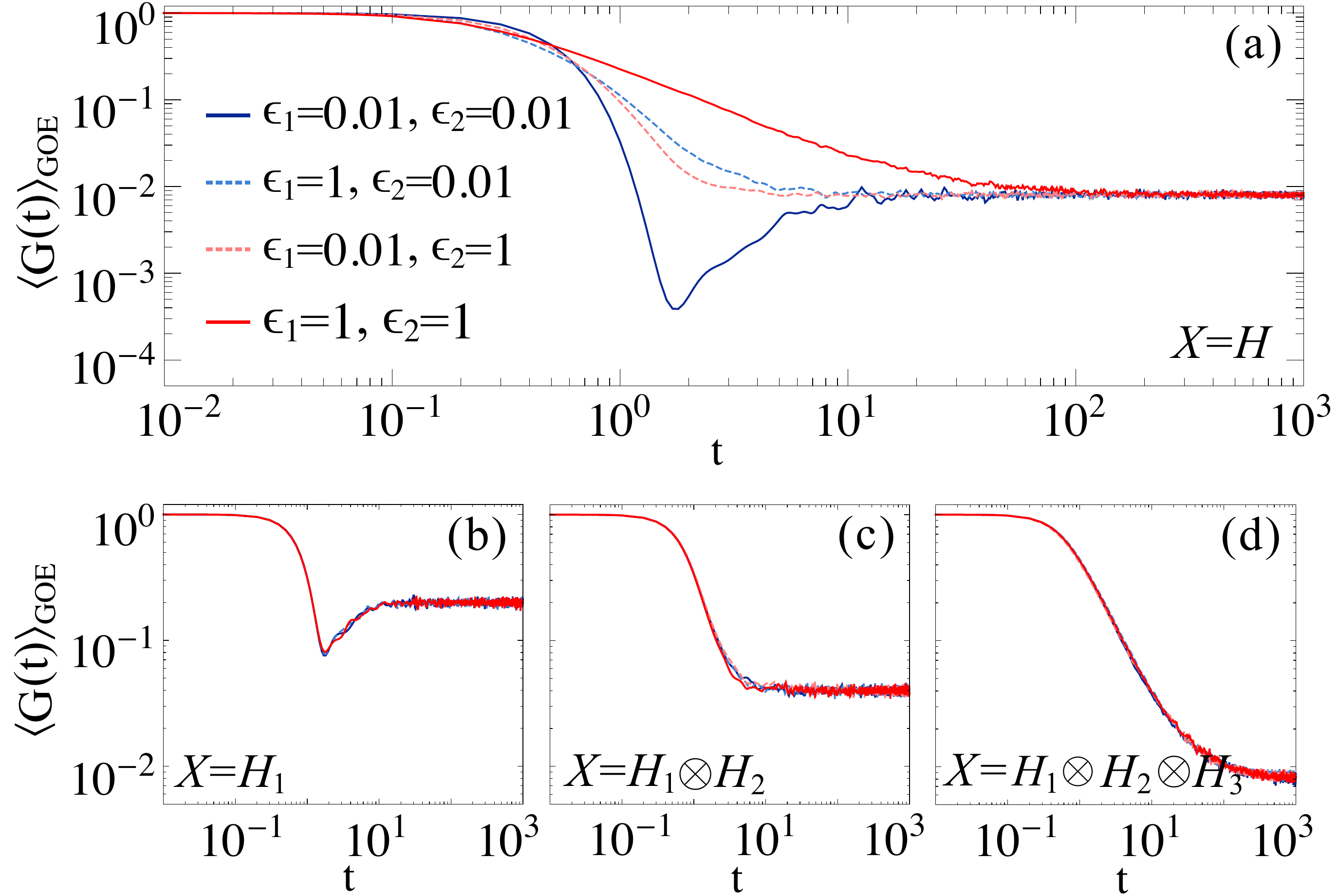}
\caption{\textbf{Chaos is suppressed by enhancing the
interactions among a tripartite system.} Equation (\protect\ref{CF})
averaged with \textrm{GOEs }for observables (a) $X=H$, (b) $X=H_{1}$, (c) $%
X=H_{1}\otimes H_{2}$, and (d) $X=H_{1}\otimes H_{2}\otimes H_{3}$,
respectively. $H_{1}$, $H_{2}$, and $H_{3}$ are sampled from \textrm{GOE}
independently, with 500 realizations, $d_{1}=d_{2}=d_{3}=5$, and $\protect%
\beta =0.01$.}
\label{fig3}
\end{figure}

Without loss of generality, we then consider a bipartite system with $%
H_{1,2}^{(0,1)}$ independently sampled from \textrm{GOEs}. As the total
system is composed of two partitions each described by a random-matrix
Hamiltonian, it is not surprising that in the absence of (or weak)
interactions the full system exhibits visible dip-ramp-plateau structure
when choosing $X=H$ [see the dark blue curve in Fig. \ref{fig2} (a)].
However, when the coupling strength $\epsilon _{1}$ between subsystems $1$
and $2$ is enhanced, the dip-ramp-plateau structure gradually washes out
[light blue curves in Fig. \ref{fig2} (a)]. 

To account for this phenomenon, we look at the characteristic function for
different choices of the observable $X=H_{1}^{(0)}$ in Fig. \ref{fig2} (b)
and $X=H_{1}^{(1)}\otimes H_{2}^{(1)}$ in Fig. \ref{fig2} (c) and show how
these choices identify the contributions to quantum chaos by the first
subsystem and the interactions between subsystems $1$ and $2$. Obviously, $%
H_{1}^{(1)}\otimes H_{2}^{(1)}$ plays an important role in diminishing the
chaotic behavior, since the characteristic function reflects no obvious ramp
structure. Indeed, from the perspective of the
nearest-neighbor level distribution, the Kronecker product of random
matrices will tend to break the Wigner-Dyson distribution under certain
conditions \cite{Tkocz2012}. When the coupling is enhanced, the interaction
term $H_{1}^{(1)}\otimes H_{2}^{(1)}$ dominates, and the chaoticity of the
whole system is gradually suppressed.

Similar phenomena exist in more structured systems, as shown in Fig. \ref%
{fig3} for a tripartite system. For simplicity, we only consider $%
H_{1,2,3}^{(0)}=$ $H_{1,2,3}^{(1)}=H_{1,2,3}^{(2)}$ and omit the superscript
therein. Both bipartite interactions (e.g., $H_{1}\otimes H_{2}$), and the
tripartite interaction $H_{1}\otimes H_{2}\otimes H_{3}$ play an import role
in decreasing the chaoticity of the composite global system, while the local
chaotic nature of each subsystem can be detected by choosing a local
observable (e.g. $X=H_{1}$) even at strong coupling.

It is worthwhile to note that the presence of interactions could also induce
chaos if the interaction tend to mix the subsystems, just as shown in the
coupled kicked rotors \cite{Lakshminarayan2016PRL}.

\section{Coupled Sachdev-Ye-Kitaev model}

The second example we consider is the coupled Sachdev-Ye-Kitaev (cSYK) model
\cite%
{Zhai2017,Song2017,Yao2017,maldacena2018eternal,Garica2019,Plugge2020,Qi2020,Sahoo2020,Haenel2021,Zhai2021}%
. A system composed of $2N$ Majorana fermions is divided into two separated
sides and each subsystem is described by the SYK model \cite%
{SachdevYe93,Kitaev15}. We consider the left and right SYK Hamiltonians $%
H_{L,R}$ with a bilinear coupling $H_{b}$. The total Hamiltonian of the cSYK
model reads

\begin{widetext}
\begin{equation}
H =H_{L}+H_{R}+\mu H_{b}=\sum_{\alpha =L,R}\sum_{1\leq k<l<m<n\leq N}J_{klmn}\chi _{k}^{\alpha}\chi _{l}^{\alpha }\chi _{m}^{\alpha }\chi _{n}^{\alpha }+\mu
i\sum_{k=1}^{N}K_{kk}\chi _{k}^{L}\chi _{k}^{R},
\end{equation}
\end{widetext}
where $\mu $ controls the strength of the the bilinear coupling and $\chi
_{k}$ denote Majorana fermion operators satisfying the anticommutation
relations $\{\chi _{k},\chi _{l}\}=\delta _{kl}$. $J_{klmn}$ and $K_{kk}$
are random coupling constants independently sampled from Gaussian
distributions with zero expectation values and $\left\langle
J_{klmn}^{2}\right\rangle =\frac{3!J^{2}}{N^{3}}$, $\left\langle
K_{kk}^{2}\right\rangle =\frac{K^{2}}{N^{2}}$. A similar model has been used
to study the holographic duality of an eternal traversable wormhole: By
preparing the SYK model in a thermofield double state and turning on the
coupling between the two sides, the wormhole is traversable in the context
of gravity \cite{maldacena2018eternal}.

\begin{figure}[t]
\centering
\includegraphics[width=0.98\linewidth]{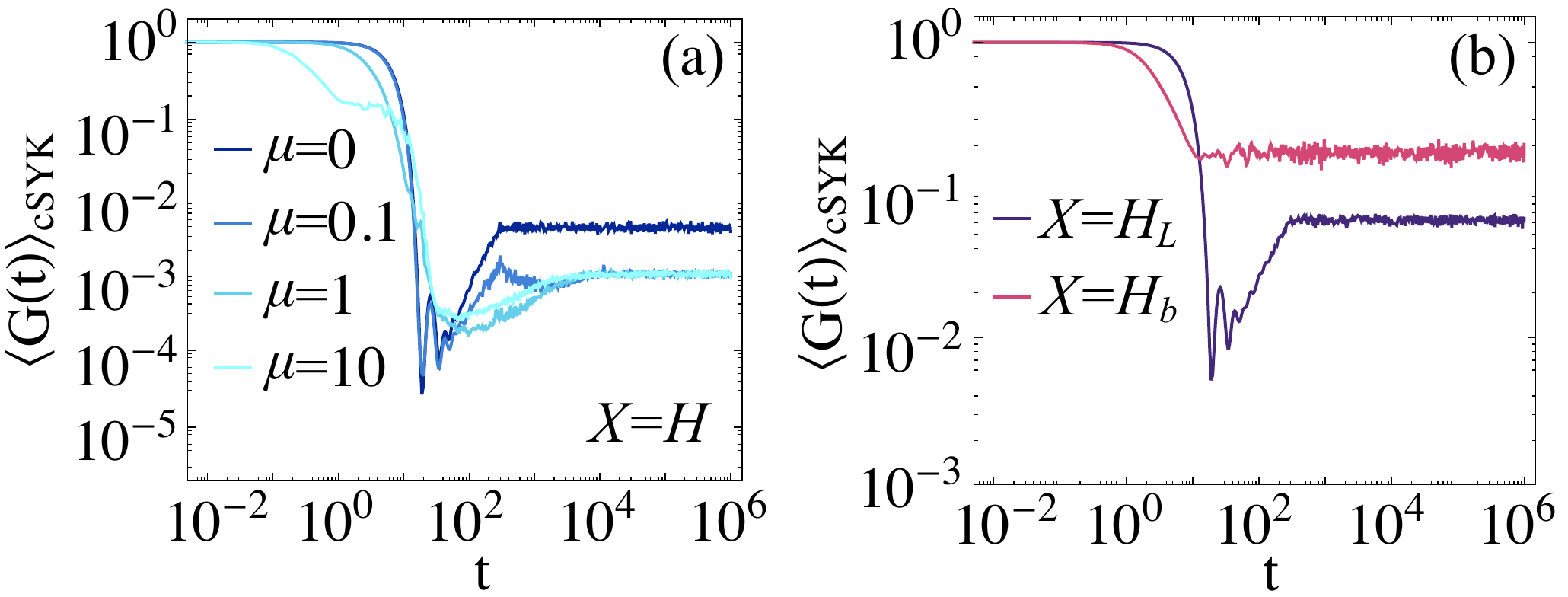}
\caption{\textbf{Chaotic behavior in a coupled Sachdev-Ye-Kitaev (cSYK)
model.} Equation (\protect\ref{CF}) averaged with cSYK\textrm{\ }for
observables (a) $X=H$, (b) $X=H_{L}$, and $X=H_{b}$, respectively. The
Hamiltonian is sampled by $500$ realizations with $2N=20$, $\protect\beta =0$%
, and $J=K=1$.}
\label{fig4}
\end{figure}

Figure \ref{fig4} shows the numerical result for the disorder-averaged $%
\langle G(t)\rangle $, in agreement with the random matrix theory.
From Fig. \ref{fig4}(b), we see that the bilinear coupling plays an
important role in controlling the chaotic behavior of the whole system, as
the generating function has no obvious dip-ramp-plateau structure. The chaotic
character of the system is robust when the bilinear coupling is weak ($\mu
\lesssim 0.1$). When enhancing the coupling, the dip-ramp-plateau structure
of the entire system is gradually washed out [Fig. \ref{fig4}(a)]. 
This implies that the entire system becomes less chaotic when
the bilinear coupling is strong, even when the chaoticity of the subsystems
remain consistent. As in the GOE example, the chaotic nature of each
subsystem can be detected by choosing a local observable (e.g. $X=H_{L}$),
as shown in Fig. \ref{fig4}(b).

\section{Discussion and conclusion}

We have introduced a protocol to directly detect quantum chaos in
interacting multipartite systems by measuring the statistical distribution
of an energy observable $X$ at thermal equilibrium. Specifically, we make
use of the absolute square value $G(t)$ of the generating function of the
eigenvalue distribution associated with the observable $X$. When the
observable equals the total Hamiltonian, $G(t)$ reduces to the SFF. For
local observables, chaotic features give rise to a dip-ramp-plateau
structure in $G(t)$, which is similar to that in SFF. $G(t)$ directly
detects the contributions to quantum chaos in a composite system from
different subsystems by choosing the observable for $k$-partite
interactions. We have shown that the coupling of chaotic systems can give
rise to the suppression of quantum chaos in the composite system, as the
interaction strength among the subsystems is increased. From the perspective of decoherence, sampling the eigenvalue statistics of a subsystem is like sampling the local energy. Quantum chaos is generally expected to be suppressed as a result of decoherence \cite{Zhenyu2021PRB,LiaoGalitski22}; see however \cite{Julien2021}.
In addition, even at strong coupling, the chaotic character of the subsystems can be unveiled by choosing $X$ as a local observable, as demonstrated by considering the
multipartite GOEs and the coupled SYK models.

Our scheme can be implemented in quantum devices, such as NMR systems \cite%
{Zhang2008,Tiago2014,Peng15} and trapped ions \cite{RN1221} by introducing
an auxiliary qubit coupled to the systems, as both the random spin and SYK
models are realizable in the laboratory \cite%
{Danshita2017,Solano2017,Pikulin2017,Jason2017,Luo2019,Babbush2019,Wei2021}.
Our approach for diagnosing the chaos in multipartite quantum systems may
thus find broad applications in interdisciplinary studies in quantum
information, quantum matter, and AdS/CFT duality, especially in analyzing
quantum chaos in structured quantum many-body systems.

\section*{Acknowledgements}

It is a pleasure to acknowledge discussions with Aurelia Chenu, Fernando J. G%
\'{o}mez-Ruiz, Mar Ferri, Wenlong You, and Yifeng Yang. This work was
supported by the National Natural Science Foundation of China under Grant
No. 12074280.


\appendix

\renewcommand{\theequation}{A\arabic{equation}}

\section{The generating function $G(t)$ averaged over ensembles}

\label{appA}

In Appendix A, we intend to briefly introduce the calculation of the
generating function $G(t)$ averaged over ensembles. The averaged $G(t)$ in
terms of annealing approximation is given by \cite{delcampo17,Zhenyu2021PRB}
\begin{equation}
\left\langle G(t)\right\rangle \doteq \frac{\left\langle \left\vert Z\left(
\beta +it\right) \right\vert ^{2}\right\rangle }{\left\langle Z(\beta
)\right\rangle ^{2}}.  \label{av-fidelity}
\end{equation}%
This annealing average is in agreement with the quenched
average $\left\langle \left\vert Z\left( \beta +it\right)
\right\vert^{2}/Z(\beta )^{2}\right\rangle $ in high temperature region
\cite{Cotler2017,Zhenyu2021PRB}. Then the denominator and numerator of Eq. (%
\ref{av-fidelity}) can be written as
\begin{equation}
\left\langle Z(\beta )\right\rangle =\int dE\rho (E)e^{-\beta E},
\label{averaged-de}
\end{equation}%
and
\begin{widetext}
\begin{equation}
\left\langle \left\vert Z\left( \beta +it\right) \right\vert
^{2}\right\rangle =\int dE\rho (E)e^{-2\beta E}+\int dEdE^{\prime }\rho
(E,E^{\prime })e^{-(\beta +it)E}e^{-(\beta -it)E^{\prime }},
\label{averaged-nu}
\end{equation}%
\end{widetext}
where $\rho (E)$ is the spectral density and $\rho (E,E^{\prime })$ is the
two-point probability density function.

\subsection{Gaussian orthogonal ensemble statistics}

For \textrm{GOE}, the spectral density and two-point probability density
function are given by \cite{Zhenyu2021PRB}
\begin{equation}
\rho (E)=\frac{1}{\sqrt{2}\sigma }\det K_{d}\left( \tilde{E},\tilde{E}%
\right) ,\text{ and }\tilde{E}:=\frac{E}{\sqrt{2}\sigma }\text{,}
\label{pdf}
\end{equation}%
and
\begin{equation}
\rho (E,E^{\prime })=\frac{1}{2\sigma ^{2}}\det \left[ \left(
\begin{array}{ll}
K_{d}(\tilde{E},\tilde{E}) & K_{d}(\tilde{E},\tilde{E}^{\prime }) \\
K_{d}(\tilde{E}^{\prime },\tilde{E}) & K_{d}(\tilde{E}^{\prime },\tilde{E}%
^{\prime })%
\end{array}%
\right) \right] ,  \label{2pdf}
\end{equation}%
respectively, and $K_{d}(x,y)$ is the kernel \cite{MethaBook}. Note that \textrm{GOE} and \textrm{GUE} share the same form of
$N$-point probability density function. The only difference is that the
kernel $K_{d}(x,y)$ in Eqs. (\ref{pdf}) and (\ref{2pdf}) for \textrm{GOE} is
a quaternion. Note that $\sigma $ is selected as $\sigma =1/\sqrt{2}$ in
Ref. \cite{MethaBook}, $\sigma =1$ in Ref. \cite{VivoBook}, and $\sigma =1/%
\sqrt{d}$ in Ref. \cite{Cotler2017}, respectively. In this paper, we keep $%
\sigma $ in the formulae for convenience.

According to Dyson's theorem \cite{MethaBook},
\begin{equation}
\det A=\mathrm{pf}\left( \mathrm{Z}_{N}\Theta \lbrack A]\right) ,
\label{Dyson}
\end{equation}
a $N\times N$ self-dual quaternion matrix $A$ can be represented by a $%
2N\times 2N$ complex matrix ($\Theta \lbrack \cdot ]$ is the matrix form of
a quaternion). Here \textrm{pf} denotes a Pfaffian and $\mathrm{Z}%
_{N}=\bigoplus_{j=1}^{N}\left(
\begin{array}{ll}
0 & 1 \\
-1 & 0%
\end{array}%
\right) _{j}$. Then Eqs. (\ref{pdf}) and (\ref{2pdf}) can be written as
\begin{equation}
\rho (E)=\frac{1}{\sqrt{2}\sigma }\mathrm{pf}\left( \mathrm{Z}_{1}\Theta
\lbrack K_{d}\left( \tilde{E},\tilde{E}\right) ]\right) ,  \label{pf-pdf}
\end{equation}%
and
\begin{equation}
\rho (E,E^{\prime })=\frac{1}{2\sigma ^{2}}\mathrm{pf}\left( \mathrm{Z}%
_{2}\Theta \left[ \left(
\begin{array}{ll}
K_{d}(\tilde{E},\tilde{E}) & K_{d}(\tilde{E},\tilde{E}^{\prime }) \\
K_{d}(\tilde{E}^{\prime },\tilde{E}) & K_{d}(\tilde{E}^{\prime },\tilde{E}%
^{\prime })%
\end{array}%
\right) \right] \right) .  \label{pf-2pdf}
\end{equation}

The above method can be straightforwardly extended to higher point
correlation functions. Similar work has been done in Ref. \cite{JunyuLiu2018}
for evaluating spectral form factors under infinite temperature.

\subsection{G(t) averaged by \textrm{GOE}s}

With Eq. (\ref{pf-pdf}), the partition function [Eq. (\ref{averaged-de})]
averaged over the \textrm{GOE}s is approximated by

\begin{equation}
\left\langle Z(x)\right\rangle _{\mathrm{GOE}}=\frac{\sqrt{d}\mathrm{I}%
_{1}(2\sigma \sqrt{d}x)}{\sigma x},  \label{pF av}
\end{equation}%
where $\mathrm{I}_{n}(\cdot )$ is the modified Bessel function of first kind
and order $n$ (Eq. (\ref{par-GOE}) in the main text).

According to Eqs. (\ref{pf-2pdf}), the imaginary time partition function
[Eq. (\ref{averaged-nu})] averaged over the \textrm{GOEs} reads
\begin{equation}
\left\langle \left\vert Z\left( \beta +it\right) \right\vert
^{2}\right\rangle _{\text{\textrm{GOE}}}=\left\langle Z(2\beta
)\right\rangle _{\mathrm{GOE}}+\left\vert \left\langle Z(\beta
+it)\right\rangle _{\mathrm{GOE}}\right\vert ^{2}+\ast ,  \label{it pF av}
\end{equation}%
where
\begin{equation}
\ast =\left\{
\begin{array}{ll}
-\frac{\alpha \sqrt{d}}{\sigma }\left[ 1-\frac{t\sigma }{\sqrt{d}}+\frac{%
t\sigma }{2\sqrt{d}}\ln \left( 1+\frac{t\sigma }{\sqrt{d}}\right) \right] ,
& t\leq 2\sqrt{d}/\sigma , \\
\frac{\alpha \sqrt{d}}{\sigma }\left[ 1-\frac{t\sigma }{2\sqrt{d}}\ln \frac{%
t+\sqrt{d}/\sigma }{t-\sqrt{d}/\sigma }\right] . & t>2\sqrt{d}/\sigma .%
\end{array}%
\right.  \label{gc4}
\end{equation}%
Considering $G(t)$ should be $1$ when $t=0$, $\alpha $ reads

\begin{equation}
\alpha \simeq \frac{\sigma }{\sqrt{d}}\left\langle Z(2\beta )\right\rangle _{%
\mathrm{GOE}}\text{.}  \label{hdiamond1}
\end{equation}%
Substitution of Eq. (\ref{hdiamond1}) and Eq. (\ref{gc4}) into Eq. (\ref{it
pF av}) together with the averaged partition function in Eq. (\ref{pF av}),
the $\mathrm{GOE}$ averaged $G(t)$ in Eq. (\ref{av-fidelity}) can be
obtained straightforwardly, see Eq. (\ref{FGOE}) and Eq. (\ref{CGOE}) in the
main text.

\begin{figure*}[t]
\centering
\includegraphics[width=0.85\linewidth]{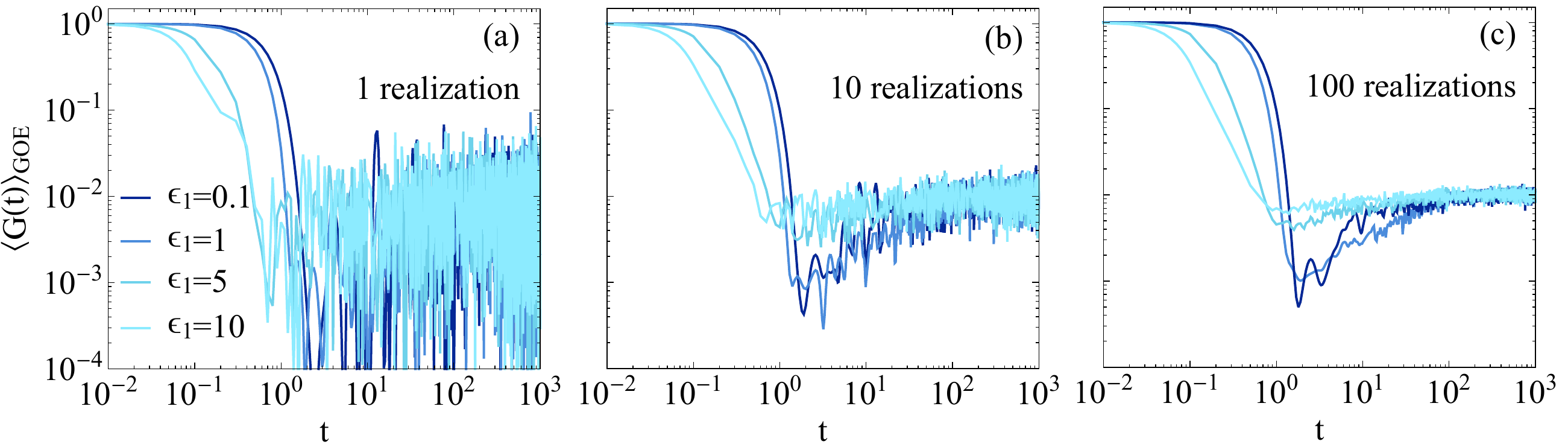}
\caption{\textbf{Different number of ensemble realizations in
Fig. \protect\ref{fig2}(a).} Data for $G(t)$ is averaged over 1, 10, and 100
realizations of GOE.}
\label{figS1}
\end{figure*}

\begin{figure*}[t]
\centering
\includegraphics[width=0.85\linewidth]{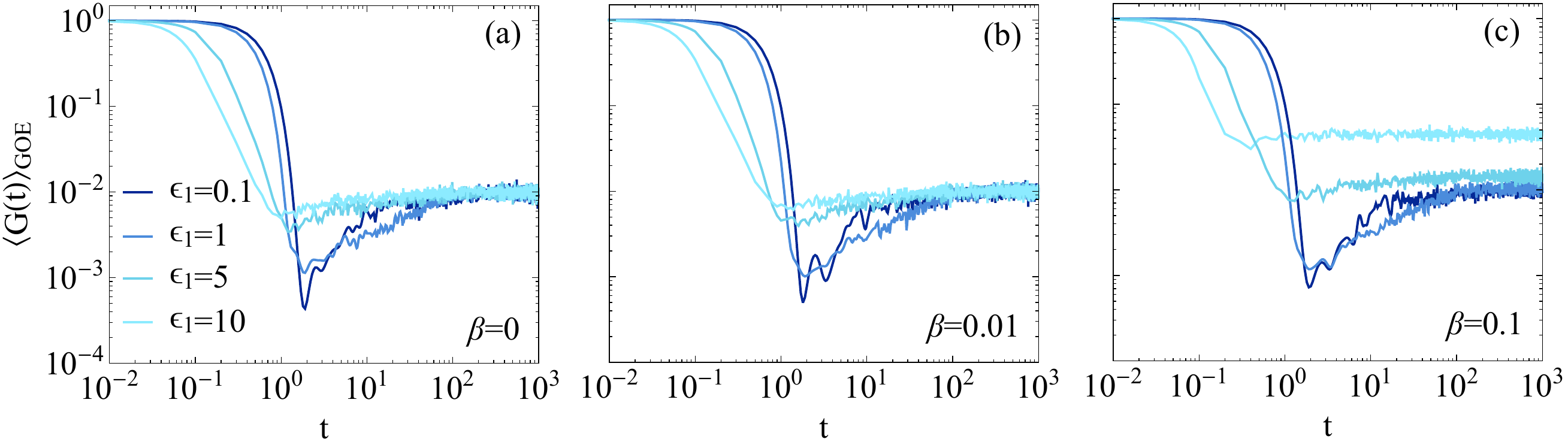}
\caption{\textbf{Temperature dependence of $G(t)$ in Fig.
\protect\ref{fig2}(a).} Data for $G(t)$ is averaged over 100 realizations of
GOE with $\protect\beta$=0, 0.01, and 0.1, respectively. }
\label{figS2}
\end{figure*}

\section{The temperature and number of realizations in ensemble average}

\label{appB} In the main text, we consider fixed number of
realizations and temperature for illustration. In this section, we append Fig. %
\ref{figS1} and Fig. \ref{figS2} to illustrate $G(t)$ with the dependence of
the number of realizations and temperature.

\bibliography{references_Chaos}

\end{document}